\documentclass[sigconf]{acmart}

\usepackage[utf8]{inputenc}
\usepackage[ngerman,main=english]{babel}
\addto\extrasenglish{\languageshorthands{ngerman}\useshorthands{"}}
\usepackage{microtype}
\usepackage{ellipsis}
\usepackage{amsmath,amsfonts}
\usepackage{booktabs}
\usepackage{paralist}
\usepackage{enumitem}
\usepackage[babel]{csquotes}
\usepackage{textcmds}
\usepackage{pdfcolfoot}
\usepackage{dblfloatfix}
\usepackage{subcaption}
\usepackage{tikz}
\usepackage{tikzpagenodes}
\usetikzlibrary{
	arrows,
	arrows.meta,
	babel,
	backgrounds,
	calc,
	fit,
	positioning,
	shapes,
	shadows,
	patterns,
}
\usepackage{tikz-layers}
\usepackage[all]{hypcap}
\newenvironment{ilenum}
	{\begin{inparaenum}[(1.)]}{\end{inparaenum}}

\newcommand{\ie}{i.\,e.\@}
\newcommand{\eg}{e.\,g.\@}

\newcommand{\p}{p.\@}

\newcommand{\wrt}{w.\,r.\,t.\@}
\newcommand{\AppSPEAR}{\textsc{AppSPEAR}}
\usepackage{url}
\makeatletter
\g@addto@macro{\UrlBreaks}{\UrlOrds}
\makeatother
\usepackage{hyperref}
\hypersetup{hidelinks,
	colorlinks=false,
	allcolors=black,
	pdfstartview=Fit,
	breaklinks=true
}
\makeatletter
\expandafter\def\expandafter\UrlBreaks\expandafter{\UrlBreaks
	\do\a\do\b\do\c\do\d\do\e\do\f\do\g\do\h\do\i\do\j%
	\do\k\do\l\do\m\do\n\do\o\do\p\do\q\do\r\do\s\do\t%
	\do\u\do\v\do\w\do\x\do\y\do\z\do\A\do\B\do\C\do\D%
	\do\E\do\F\do\G\do\H\do\I\do\J\do\K\do\L\do\M\do\N%
	\do\O\do\P\do\Q\do\R\do\S\do\T\do\U\do\V\do\W\do\X%
	\do\Y\do\Z\do\*\do\-\do\~\do\'\do\"\do\-}%
\makeatother

\setcopyright{none}
\renewcommand\footnotetextcopyrightpermission[1]{}

\begin{document}

\pagestyle{empty}

\title{Trusted Enforcement of Application-specific Security Policies}

\author{Marius Schlegel}
\orcid{0000-0001-6596-2823}
\affiliation{%
	\institution{TU Ilmenau, Germany}
}
\email{marius.schlegel@tu-ilmenau.de}

\renewcommand{\shortauthors}{M. Schlegel}

\begin{abstract}
	While there have been approaches for integrating security policies into operating systems (OSs) for more than two decades, applications often use objects of higher abstraction requiring individual security policies with application-specific semantics. Due to insufficient OS support, current approaches for enforcing application-level policies typically lead to large and complex trusted computing bases rendering tamperproofness and correctness difficult to achieve. To mitigate this problem, we propose the application-level policy enforcement architecture \AppSPEAR{} and a C++ framework for its implementation. The configurable framework enables developers to balance enforcement rigor and costs imposed by different implementation alternatives and thus to easily tailor an \AppSPEAR{} implementation to individual application requirements. We especially argue that hardware-based trusted execution environments offer an optimal balance between effectiveness and efficiency of policy protection and enforcement. This claim is substantiated by a practical evaluation based on an electronic medical record system.
\end{abstract}


\maketitle


\section{Introduction}

The rigorous and correct enforcement of application"=specific security goals in today's complex software systems is far from being an easy task. In order to protect the confidentiality and integrity of security-""critical resources (\eg{} customer information or accounting information), such systems increasingly rely on a \emph{security policy}: an automatically enforced set of rules that control evaluating, granting, and revoking access privileges \cite{Ferraiolo07a,Servos17a}, control tracing, classifying, and confining potential flows of information \cite{Bell76a,Zeldovich08a}, or control isolating domains of users and resources \cite{Goguen82a,Rushby92a}.

For almost two decades the idea has been pursued to integrate security policies directly into operating systems (OSs) \cite{Wat13}. The control of operations on OS objects (\ie{} resources described by abstractions such as files, processes, sockets, etc.) is made possible by the extension of standard OS kernels (\eg{} Linux, FreeBSD, Android) by mandatory access control (AC) and information flow control mechanisms \cite{LS01a,VW03,KYB+07,KT09,BHS13,SC13,SMS14}.

At application level, for instance, enterprise resource planning (ERP) systems typically use role-based AC policies that reflect a company's organizational structure \cite{AZK+11,BGB+05}, workflow management systems use information flow policies \cite{CGW16,WL10}, database systems use label-based AC policies to control access to relations and views \cite{Oraclec,IBMb}, (social) information systems use relationship-based AC policies on user data \cite{RFC+15,Fon11}, and Big Data and IoT platforms rely on attribute-based AC policies \cite{GPS18,BHR18}. Compared to OSs, the objects used by applications are typically subject of security policies on a higher, application-specific abstraction level. This requires policy rules with application-specific semantics which typically differ significantly from those at OS level and which are also specific for each individual application.

Contemporary OSs do not adequately support application-""specific security policies. For this reason, developers often integrate security policies directly into applications which results in large and heterogeneous trusted computing base (TCB) implementations of applications. Due to the close integration of security-relevant functionality and application logic, the precise identification of an application's TCB perimeter is hard if not impossible, rendering correctness properties difficult to achieve.

To enable precisely defined application TCBs as well as a rigorous and trusted enforcement of individual application security policies, this paper argues for an alternative approach. First, the foundation is laid by strictly separating security-relevant functionality from (potentially untrusted) application logic. Second, this separation is implemented by an isolation mechanism. Adjusting the strength of isolation and the costs for crossing isolation boundaries, the approach is open for different isolation mechanisms ranging from language-based (\eg{} type"=safe programming languages and compilers) and OS-based mechanisms (\eg{} virtual address spaces via processes) to trusted execution technologies (\eg{} Intel SGX) \cite{SWG+16}.

In particular, Intel SGX provides trusted execution environments (TEEs), so-called \emph{enclaves}, which isolate security-sensitive parts of application code and data from the OS kernel, hypervisor, BIOS, and other applications using a hardware-protected memory region \cite{MAB+13,Int19b}. While this significantly reduces the size of application TCB implementations, crossing isolation boundaries typically imposes high costs compared to conventional isolation mechanisms \cite{WBA17,OLM+17,TZY+18}. Therefore, we examine its suitability for our approach.

Specifically, we make the following contributions:
\begin{ilenum}
	\item We introduce \AppSPEAR{}~-- an application"=level security policy enforcement architecture providing a functional framework for implementing the reference monitor principles (§\,\ref{sec:archdesign}).
	\item Balancing rigor of policy enforcement (\emph{effectiveness}) and isolation/""communication costs (\emph{efficiency}), we discuss alternatives for implementing \AppSPEAR{} and the consequences of using different isolation mechanisms (§\,\ref{sec:archimpl:balancing}). We have casted this spectrum of implementation alternatives into a software framework, which highlights the developer support provided. Moreover, we describe our experiences with the integration of \AppSPEAR{} into the electronic medical record system OpenMRS (§\,\ref{sec:archimpl:study}).
	\item We present an evaluation of the \AppSPEAR{} framework based on our implementation showing the practical runtime costs imposed by different implementation alternatives (§\,\ref{sec:eval}).
\end{ilenum}


\section{Architecture Design}
\label{sec:archdesign}

This section introduces \AppSPEAR{}. The purpose of the architecture is to provide a functional framework for the rigorous enforcement of application"=specific security policies. In the following, we first discuss fundamental security policy and architecture design principles (§\,\ref{sec:archdesign:req}). Based on those, we describe our architecture, including its components, their tasks and interrelationships (§\,\ref{sec:archdesign:archdesign}).

\subsection{Requirements and Design Principles}
\label{sec:archdesign:req}

\paragraph*{Application-specific Security Policies.}
Security policies are usually well tailored to their respective application domain. While differ semantically among different classes of policies, namely AC \cite{Ferraiolo07a,Servos17a}, information flow \cite{Bell76a,Zeldovich08a}, and noninterference \cite{Goguen82a,Rushby92a}, they all aim at guarantees towards security properties.

As one of the most prominent security policy classes in practice, we focus on AC policies for the scope of this paper. Furthermore, to support a broad range of different model abstractions (\eg{} roles, labels, contexts, risks, relationships, or attributes in general \cite{Ferraiolo07a,Biswas16a,Shebaro13a,Chen12a,Fon11,Jin12a}), we consider an AC policy from the perspective of the enforcement mechanisms as a black box of rules. These rules authorize any operation $\mathit{op} \in \mathit{OP}$ (\eg{} \emph{read} from or \emph{append} to electronic patient record (EPR) objects) related to a vector of entities $e = (e_i)_{i = 1}^n \in E^n$ (\eg{} application users, EPRs or documents). The interface of an AC policy $P$ can be defined as a semantically neutral \emph{access control function} (ACF)
\begin{equation*}
	\label{eq:1}
	f_P : E^n \times \mathit{OP} \rightarrow \{\text{true},\allowbreak \text{false}\}.
\end{equation*}
For example within an EPR management system, appending new checkup results to a patient's EPR could then by requested by $f_P(\langle{}\text{Alice}, \text{EPR}_\text{Bob}\rangle{}, \text{append})$.

In recent years, research rendered a number of modeling schemes for context-aware AC policies that model adaptive authorization decisions based on physical or logical features, represented by local context variables. Values of such variables may be either computed, \eg{} time, date, and resource usage, or perceived by sensors, \eg{} temperature, geographic location, and NFC device proximity \cite{Zhang10a,Tripunitara15a,Hsu16a,Shebaro13a}. By considering a vector of context values $v = (v_j)_{j = 1}^m \in V^m$ (\eg{} time or geolocation) additionally to $f_P$, the interface of a context-aware AC policy $P'$ can be defined by an ACF
\begin{equation*}
	\label{eq:2}
	f_{P'} : E^n \times V^m \times \mathit{OP} \rightarrow \{\text{true},\allowbreak \text{false}\}.
\end{equation*}
These context values may be input for a risk evaluation metric (as typically used in risk-based AC models) which may calculate a resulting numerical value~\cite{Ni10a}. In the simplest case, this value is compared with a risk-threshold to assess the risk as either acceptable (permitting the original access request) or unacceptable (denying the original request).

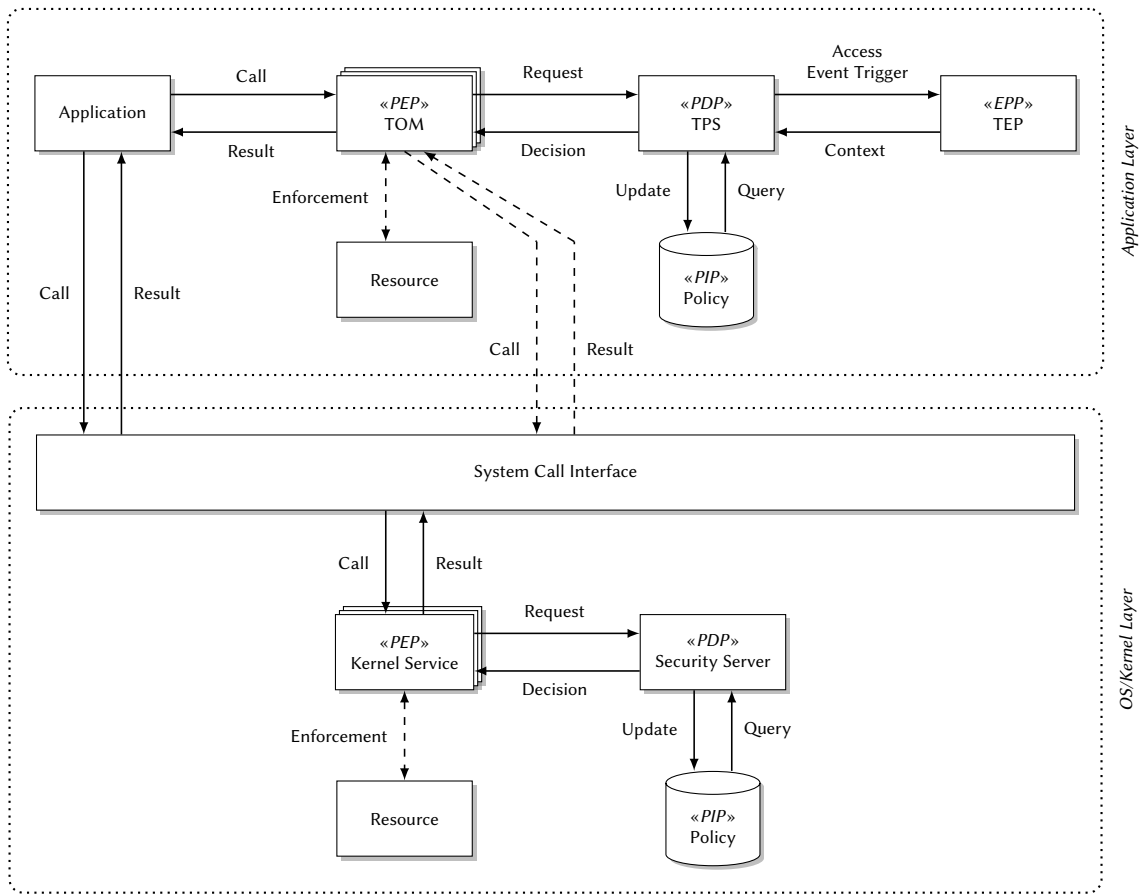
\begin{figure*}[!b]
	\centering
	\begin{tikzpicture}
		\def\nodeminheight{10mm}
		\def\nodeminwidth{18mm}
		\def\nodedistvert{12mm}
		\def\nodedisthor{22mm}
		\def\scishift{-10pt}
		\tikzstyle{every node}=[
		node font=\sffamily\footnotesize,
		minimum height=\nodeminheight,
		minimum width=\nodeminwidth,
		inner sep=2mm,
		align=center,
		node distance=\nodedistvert and \nodedisthor,
		]
		\tikzstyle{comp}=[
		fill=white,
		drop shadow,
		draw,
		]
		\tikzstyle{edge label}=[
		inner sep=1.5mm,
		minimum width=0mm,
		minimum height=0mm,
		]
		\def\size{4*\nodeminwidth+3*\nodedisthor}
		\def\shiftdist{.25cm}
		\def\appresNodeShift{-20mm+3*\shiftdist}
		\node[comp] (tom) [anchor=base,
			general shadow={
					shadow scale=1,
					shadow xshift=1ex,
					shadow yshift=1ex,
					drop shadow={
							shadow xshift=1.5ex,
							shadow yshift=.5ex,
						},
					drop shadow={
							shadow xshift=1ex,
							shadow yshift=0ex,
						},
					draw,
					fill=white,
				},
			general shadow={
					shadow scale=1,
					shadow xshift=.5ex,
					shadow yshift=.5ex,
					draw,
					fill=white,
				}
		] {\guillemotleft\textit{PEP}\guillemotright\\TOM};
		\node[comp] (appres) [below=of tom] {Resource};
		\node[comp] (app) [left=of tom] {Application};
		\node[comp] (tps) [right=of tom] {\guillemotleft\textit{PDP}\guillemotright\\TPS};
		\node[comp] (apppol) [shape=cylinder,aspect=0.3,shape border rotate=90,minimum width=1.25cm,anchor=shape center] at (appres-|tps) {\guillemotleft\textit{PIP}\guillemotright\\Policy};
		\node[comp] (tep) [right=of tps] {\guillemotleft\textit{EPP}\guillemotright\\TEP};
		\coordinate (virtnode-between-appres-apppol) at ($(tom.south)!.5!(tps.south)$);
		\node (virtnode-above-sci) [below=of virtnode-between-appres-apppol] {};
		\node[comp] (sci) [below=of virtnode-above-sci,minimum width=\size,yshift=\scishift] {System Call Interface};
		\node (virtnode1) [below=of tom] {};
		\node (virtnode2) [below=of virtnode1] {};
		\node[comp] (kernservice) [below=of virtnode2,yshift=1.5*\scishift,
			general shadow={
					shadow scale=1,
					shadow xshift=1ex,
					shadow yshift=1ex,
					drop shadow={
							shadow xshift=1.5ex,
							shadow yshift=.5ex,
						},
					drop shadow={
							shadow xshift=1ex,
							shadow yshift=0ex,
						},
					draw,
					fill=white,
				},
			general shadow={
					shadow scale=1,
					shadow xshift=.5ex,
					shadow yshift=.5ex,
					draw,
					fill=white,
				}
		] {\guillemotleft\textit{PEP}\guillemotright\\Kernel Service};
		\node[comp] (ss) [right=of kernservice] {\guillemotleft\textit{PDP}\guillemotright\\Security Server};
		\node[comp] (kernres) [below=of kernservice] {Resource};
		\node[comp] (kernpol) [shape=cylinder,aspect=0.3,shape border rotate=90,minimum width=1.25cm,anchor=shape center] at (kernres-|ss) {\guillemotleft\textit{PIP}\guillemotright\\Policy};
		\node (kernepp) [left=of ss] {};
		\draw[
			->,
			>=latex,
			semithick,
		]
		([xshift=-\shiftdist]app.south) edge node [edge label,left,xshift=1pt] {Call} ([xshift=-\shiftdist]sci.north -| app.south)
		([xshift=+\shiftdist]sci.north -| app.south) edge node [edge label,right] {Result} ([xshift=+\shiftdist]app.south)
		([xshift=-1*\shiftdist]tom.south) edge [dashed,<->] node [edge label,left] {Enforcement} ([xshift=-1*\shiftdist]appres.north)
		([xshift=0*\shiftdist]tom.south) edge [-,dashed] ([xshift=-\shiftdist]virtnode-above-sci.north)
		([xshift=-\shiftdist]virtnode-above-sci.north) edge [dashed] node [edge label,left,yshift=-4pt] (label-tomtosci) {Call} ([xshift=-\shiftdist]sci.north)
		([xshift=\shiftdist]sci.north) edge [-,dashed] node [edge label,right,yshift=-4pt]{Result} ([xshift=+\shiftdist]virtnode-above-sci.north)
		([xshift=+\shiftdist]virtnode-above-sci.north) edge [dashed] ([xshift=1*\shiftdist]tom.south)
		([xshift=-\shiftdist]tps.south) edge node [edge label,left] {Update} ([xshift=-\shiftdist]apppol.north)
		([xshift=+\shiftdist]apppol.north) edge node [edge label,right] {Query} ([xshift=+\shiftdist]tps.south)
		([yshift=+\shiftdist]app.east) edge node [edge label,above] {Call} ([yshift=+\shiftdist]tom.west)
		([yshift=-\shiftdist]tom.west) edge node [edge label,below] {Result} ([yshift=-\shiftdist]app.east)
		([yshift=+\shiftdist]tom.east) edge node [edge label,above] {Request} ([yshift=+\shiftdist]tps.west)
		([yshift=-\shiftdist]tps.west) edge node [edge label,below] {Decision} ([yshift=-\shiftdist]tom.east)
		([xshift=-\shiftdist]sci.south -| kernservice.north) edge node [edge label,left] {Call} ([xshift=-\shiftdist]kernservice.north)
		([xshift=+\shiftdist]kernservice.north) edge node [edge label,right] {Result} ([xshift=+\shiftdist]sci.south -| kernservice.north)
		([yshift=+\shiftdist]tps.east) edge node [edge label,above] (label-tpstotep) {Access\\Event Trigger} ([yshift=+\shiftdist]tep.west)
		([yshift=-\shiftdist]tep.west) edge node [edge label,below] {Context} ([yshift=-\shiftdist]tps.east)
		(kernservice.south) edge [dashed,<->] node [edge label,left] {Enforcement} (kernres.north)
		([xshift=-\shiftdist]ss.south) edge node [edge label,left] {Update} ([xshift=-\shiftdist]kernpol.north)
		([xshift=+\shiftdist]kernpol.north) edge node [edge label,right] {Query} ([xshift=+\shiftdist]ss.south)
		([yshift=+\shiftdist]kernservice.east) edge node [edge label,above] {Request} ([yshift=+\shiftdist]ss.west)
		([yshift=-\shiftdist]ss.west) -- node [edge label,below] {Decision} ([yshift=-\shiftdist]kernservice.east)
		;
		\tikzstyle{outbox}=[
		shape=rectangle,
		draw,
		dotted,
		thick,
		rounded corners=1.8mm,
		inner sep=10pt,
		]
		\begin{scope}[on behind layer]
			\node[fit={(app)(tom)(appres)(tps)(apppol)(tep)($(label-tpstotep.north)+(0, -1pt)$)($(label-tomtosci)$)},outbox, label={[name=l,rotate=90,anchor=south,yshift=-24pt, font=\itshape]right:Application Layer}] (applayer) {};
			\node[fit=(sci)(kernservice)(kernres)(ss)(kernpol)(kernepp),outbox,label={[name=l,rotate=90,anchor=south,yshift=-24pt,font=\itshape]right:OS/Kernel Layer}] (oslayer) {};
		\end{scope}
	\end{tikzpicture}

	\caption{%
		Functional architecture for mutually independent security policy enforcement on application layer (based on PM/""NGAC) and OS layer (based on SELinux/""Flask).}
	\label{fig:secarch}
	\Description[<short description>]{<long description>}
\end{figure*}

\paragraph*{Reference Monitor Principles.}
In general, security policies are part of a system's TCB. Subsequently, we consider a TCB from a functional perspective as the set of all functions that are necessary and sufficient for implementing a system's security properties. The part of a software architecture that implements the TCB forms the \emph{security architecture}.

For more than four decades, the reference monitor principles \cite{And72} have provided fundamental guidelines for the design and implementation of security architectures. These principles include three rules requiring a security architecture core (referred to as \emph{reference monitor}) that is (1.)~inevitably involved in any security-related interaction (RM~1, \emph{total mediation}), (2.)~protected from unauthorized manipulation (RM~2, \emph{tamperproofness}), and (3.)~as small and simple as possible in terms of functional complexity and amount of code (RM~3, \emph{verifiability}).

According to RM~1, it must be guaranteed that any actions on security-relevant objects are inevitably controlled by the policy. In case of security-policy-controlled OSs, this is achieved by calling the security policy within the OS kernel services (\eg{} process management, filesystem, or I/O subsystem) before the actual execution of any action (\eg{} \emph{fork}, \emph{read}, or \emph{send}) on an OS object (\ie{} a resource described by abstractions such as processes, files, or sockets)~\cite{SVS01}. At application level, RM~1 can be achieved by the separation of any functions relevant to the policy enforcement and the application object management, on the one hand, from functions solely responsible for the application logic, on the other hand. Consequently, any security-relevant object and any function for accessing such an object belongs to the TCB. If an application object is accessed by an application logic function, an immediate entry into the TCB takes place. Accordingly, by calling the security policy before any security-relevant object access within the TCB boundaries, the policy cannot be bypassed.

Moreover, the precise delineation of the TCB paves the way for protecting the integrity of an application security policy from unauthorized manipulation (cf. RM~2). The functional separation is implemented by isolation mechanisms~\cite{SWG+16}, which, depending on the assumed attacker model and the degree of tamperproofness required, range from language-based isolation (\eg{} policy execution in a class instance separated by type-""safety-""checking compilers) to OS-""based isolation (\eg{} policy execution in a separate virtual address space via a process) and virtualization techniques (\eg{} policy execution in a virtual machine), as well as hardware-""based
isolation (\eg{} policy execution in an SGX enclave) to total physical separation (\eg{} policy execution on a dedicated server).

By adapting RM~3, it generally holds that the smaller a TCB's functional perimeter and the lower its complexity, the better a TCB can be analyzed regarding correctness properties. A precise functional perimeter reduces TCB size and complexity because the functions solely associated with the application logic are no longer part of the TCB. Furthermore, to facilitate the verifiability of certain parts of the TCB, in particular security policies \wrt{} model properties such as dynamic state reachability in dynamic AC models \cite{SYG+11,Tripunitara13a,Schlegel20a,Schlegel21a}, an additional separation within the TCB boundaries of policy-specific from policy-independent TCB functions should be anchored within the architecture.

\subsection{\AppSPEAR{} Design}
\label{sec:archdesign:archdesign}

An important representative of security architectures that implements the reference monitor principles is the Flask security architecture~\cite{SSL+99}. Motivated by enabling the flexible interchangeability of security policies, the policy logic, referred to as \emph{policy decision point} (PDP), is separated from the policy enforcement, distributed over \emph{policy enforcement points} (PEPs). This key idea has been adopted by SELinux \cite{LS01a}, extending the Linux kernel by mandatory AC, and followed by a wide range of OSs \cite{Wat13}. The SELinux architecture comprises PEPs located in the OS kernel services (referred to as \emph{object managers}) and a singular PDP (referred to as \emph{security server}) providing an OS security policy runtime environment (see Fig.~\ref{fig:secarch}).

Comparing Flask's/""SELinux' objective of a rigorous and flexible policy enforcement on OS level with ours for the application level, we argue to adapt these principles. The general idea is to instantiate Flask/""SELinux-""alike architecture components for each policy-""controlled application which \emph{complement} the architecture (and security policy) at kernel level. This idea is illustrated in Fig.~\ref{fig:secarch}. Moreover, we consider state of the art in the enforcement of context-/attribute-""based AC policies as represented by the Policy Machine \cite{FGJ15} and its successor, Next-""Generation Access Control (NGAC) \cite{Ferraiolo16a}.\footnote{Since we focus on policy enforcement, we omit a \emph{policy administration point} (PAP) as entry point for administration.}

\paragraph*{Trusted Object Managers.}
In order to implement our architecture at application level according to the discussed design principles, in general, a separation between security-""relevant functionality and security-""irrelevant application logic is required. \emph{Trusted object managers} (TOMs) are architecture components to rigorously isolate trusted and non-trusted application parts and comprise all functions which are required for establishing authenticity, integrity, and confidentiality of application objects.

Consequently, a TOM provides an encapsulated object access interface for actions such as reading or modifying security-relevant application objects (\eg{} patient record objects or customer record objects). This interface is implemented by local or remote procedures or methods (depending on the strength of security isolation applied). Inspired by object-oriented software design, a generic TOM provides a generic object abstraction and abstract basic functions for its management, such as create, read, write, and destroy. For a concrete application, type-specific TOMs have to be derived from that generic TOM, which implement the abstract object and functions according to the type of object. Hence, a separate TOM is implemented for each object type analogous to the OS kernel services acting OS object managers.

When an application requests to perform a particular action on a TOM-""managed object, the corresponding application security policy is immediately involved through the interface between a TOM and the \emph{trusted policy server} (TPS). In this way, (1.) any security-""relevant subject-object interaction is inevitably controlled by the security policy and (2.) decisions can be enforced tamperproof, because the TOMs are part of the TCB and, thus, are isolated from the untrusted application logic. Finally, depending on the granted or denied access, the corresponding output of the object access is returned to the calling application logic function.

For many applications it is practically inevitable to use OS objects in addition to the application-specific ones. Actions on OS objects are controlled by the OS security policy, if present. Scattering application policy rules on both OS and application layers would contradict the TCB properties we aim for. Furthermore, if there is no OS policy present, such accesses would not be controlled at all. To maintain the responsibility of application policies here and to not require an OS policy for each object type an application uses, a simple wrapper for OS objects in form of a separate TOM is established. To reduce development effort, these may be reused (or generated automatically) based on a reference implementation. These TOMs may also be used by other TOMs to control accesses to OS objects and system calls at application~level.

In order to ensure the authenticity of application objects appearing as policy entities, TOMs are responsible for managing entity identifications (comparable with SELinux security identifiers) and for assigning them correctly as well as irrevocably to their implementations as runtime or persistent application objects such that they can be clearly identified at any time.

If we isolate architectural components from each other according to the design principles discussed in §\,\ref{sec:archdesign:req}, the communication between these components requires the crossing of isolation boundaries, which in turn causes certain costs depending on the type of isolation (\eg{} IPC when isolating via virtual address spaces). To reduce those costs, for instance for the communication of TOMs and the TPS, a caching mechanism may be used for policy decisions, similar to the SELinux access vector cache. However, it should be noted in particular that when implementing a stateful application policy, it may be necessary to explicitly manage cache consistency by invalidating cached access decisions which may become invalid due to policy state changes.

\paragraph*{Trusted Policy Server.}
The TPS represents \AppSPEAR{}'s PDP. Its main task is to provide a policy runtime environment including data structures that represent policy abstractions and components, authorization rules, and state transition scheme and policy states (the last two in case of a dynamic policy). To keep the TPS flexible enough to accommodate policies based on different modeling schemes, it does not make sense to limit the policy implementation within the TPS to a specific model. In contrast, providing the TPS with a wide range of models would lead to a universal range of functions, which is not consistent with the principle of a small functional TCB perimeter and the verifiability principle.

Here, the approach is to support exclusively the present policy and to provide only functions necessary for that policy. Therefore, a policy implementation is specifically tailored to its application while maintaining the interfaces for TOM-to-TPS and TPS-to-TOM communication. In order to reduce the implementation effort, developer support is provided in the form of TPS libraries implementing the data structures and functionality of frequently used security models. Beyond that, we aim to extend developer support even further: By providing a convenient policy specification language \cite{Amthor20a} and a compiler to generate policy-specific TPSs, the effort is reduced to the specification of policy rules.

Finally, to deal with application and system failures, crashes, or reboots, the TPS comprises functions for the persistent and secure storage of the policy's state. Depending on the required guarantees, different strategies (\eg{} logging diffs, complete state backups, etc.) are applied at different points in time (\eg{} fixed times, after each state change, etc.).

\paragraph*{Trusted Event Processor.}
The \emph{trusted event processor} (TEP) is an optional add-on to provide context-based security policies information about the physical and logical context of a system according to their interface (cf.~$f_{P'}$, §\,\ref{sec:archdesign:req}). The TEP may be implemented in a reactive manner based on asynchronous triggers instead of regular function calls. These which may originate from local hard- and software components, \eg{} GPS, clock, or temperature~sensors, typically implemented by an OS interrupt mechanism. Any access decision of the TPS may also trigger an event which is needed for logging and auditing.


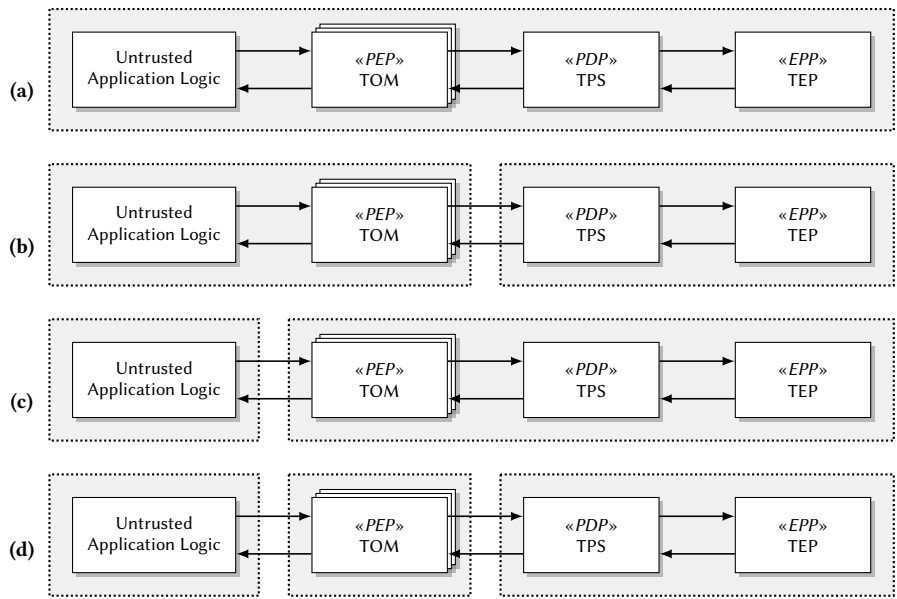
\begin{figure*}[!b]
	\def\nodeminheight{10mm}
	\def\nodeminwidth{18mm}
	\def\nodedistvert{12mm}
	\def\nodedisthor{10mm}
	\def\boxboarder{.3}
	\def\boxfillcolor{black!6}
	\def\spacebetweenvariants{0pt}
	\begin{subfigure}[!tb]{.85\textwidth}
		\centering
		\begin{minipage}[c]{.025\textwidth}
			\subcaption{}
			\label{fig:impl_a}
		\end{minipage}
		~%
		\begin{minipage}[c]{.65\textwidth}
			\begin{tikzpicture}
				\tikzstyle{every node}=[
				node font=\sffamily\footnotesize,
				minimum height=\nodeminheight,
				minimum width=\nodeminwidth,
				inner sep=2mm,
				align=center,
				node distance=\nodedistvert and \nodedisthor,
				]
				\tikzstyle{comp}=[
				fill=white,
				drop shadow,
				draw,
				]
				\def\size{4*\nodeminwidth+3*\nodedisthor}
				\def\shiftdist{.25cm}
				\def\appresNodeShift{-20mm+3*\shiftdist}
				\node[comp] (tom) [anchor=base,
					general shadow={
							shadow scale=1,
							shadow xshift=1ex,
							shadow yshift=1ex,
							drop shadow={
									shadow xshift=1.5ex,
									shadow yshift=.5ex,
								},
							drop shadow={
									shadow xshift=1ex,
									shadow yshift=0ex,
								},
							draw,
							fill=white,
						},
					general shadow={
							shadow scale=1,
							shadow xshift=.5ex,
							shadow yshift=.5ex,
							draw,
							fill=white,
						}
				] {\guillemotleft\textit{PEP}\guillemotright\\TOM};
				\node[comp] (app) [left=of tom] {Untrusted\\Application Logic};
				\node[comp] (tps) [right=of tom] {\guillemotleft\textit{PDP}\guillemotright\\TPS};
				\node[comp] (tep) [right=of tps] {\guillemotleft\textit{EPP}\guillemotright\\TEP};
				\draw[
					->,
					>=latex,
					semithick,
				]
				([yshift=+\shiftdist]app.east) edge ([yshift=+\shiftdist]tom.west)
				([yshift=-\shiftdist]tom.west) edge ([yshift=-\shiftdist]app.east)
				([yshift=+\shiftdist]tom.east) edge ([yshift=+\shiftdist]tps.west)
				([yshift=-\shiftdist]tps.west) edge ([yshift=-\shiftdist]tom.east)
				([yshift=+\shiftdist]tps.east) edge ([yshift=+\shiftdist]tep.west)
				([yshift=-\shiftdist]tep.west) -- ([yshift=-\shiftdist]tps.east)
				;
				\begin{scope}[on background layer]
					\draw[thick,densely dotted,fill=\boxfillcolor] ($(app.north west)+(-\boxboarder,\boxboarder)$) rectangle ($(tep.south east)+(\boxboarder,-\boxboarder)$);
				\end{scope}
			\end{tikzpicture}
			\vspace{\spacebetweenvariants}
		\end{minipage}
	\end{subfigure}
	\begin{subfigure}[!tb]{.85\textwidth}
		\centering
		\begin{minipage}[c]{.025\textwidth}
			\subcaption{}
			\label{fig:impl_b}
		\end{minipage}
		~%
		\begin{minipage}[c]{.65\textwidth}
			\begin{tikzpicture}
				\tikzstyle{every node}=[
				node font=\sffamily\footnotesize,
				minimum height=\nodeminheight,
				minimum width=\nodeminwidth,
				inner sep=2mm,
				align=center,
				node distance=\nodedistvert and \nodedisthor,
				]
				\tikzstyle{comp}=[
				fill=white,
				drop shadow,
				draw,
				]
				\def\size{4*\nodeminwidth+3*\nodedisthor}
				\def\shiftdist{.25cm}
				\def\appresNodeShift{-20mm+3*\shiftdist}
				\node[comp] (tom) [anchor=base,
					general shadow={
							shadow scale=1,
							shadow xshift=1ex,
							shadow yshift=1ex,
							drop shadow={
									shadow xshift=1.5ex,
									shadow yshift=.5ex,
								},
							drop shadow={
									shadow xshift=1ex,
									shadow yshift=0ex,
								},
							draw,
							fill=white,
						},
					general shadow={
							shadow scale=1,
							shadow xshift=.5ex,
							shadow yshift=.5ex,
							draw,
							fill=white,
						}
				] {\guillemotleft\textit{PEP}\guillemotright\\TOM};
				\node[comp] (app) [left=of tom] {Untrusted\\Application Logic};
				\node[comp] (tps) [right=of tom] {\guillemotleft\textit{PDP}\guillemotright\\TPS};
				\node[comp] (tep) [right=of tps] {\guillemotleft\textit{EPP}\guillemotright\\TEP};
				\draw[
					->,
					>=latex,
					semithick,
				]
				([yshift=+\shiftdist]app.east) edge ([yshift=+\shiftdist]tom.west)
				([yshift=-\shiftdist]tom.west) edge ([yshift=-\shiftdist]app.east)
				([yshift=+\shiftdist]tom.east) edge ([yshift=+\shiftdist]tps.west)
				([yshift=-\shiftdist]tps.west) edge ([yshift=-\shiftdist]tom.east)
				([yshift=+\shiftdist]tps.east) edge ([yshift=+\shiftdist]tep.west)
				([yshift=-\shiftdist]tep.west) -- ([yshift=-\shiftdist]tps.east)
				;
				\begin{scope}[on background layer]
					\draw[thick,densely dotted,fill=\boxfillcolor] ($(app.north west)+(-\boxboarder,\boxboarder)$) rectangle ($(tom.south east)+(\boxboarder,-\boxboarder)$);
					\draw[thick,densely dotted,fill=\boxfillcolor] ($(tps.north west)+(-\boxboarder,\boxboarder)$) rectangle ($(tep.south east)+(\boxboarder,-\boxboarder)$);
				\end{scope}
			\end{tikzpicture}
			\vspace{\spacebetweenvariants}
		\end{minipage}
	\end{subfigure}
	\begin{subfigure}[!tb]{.85\textwidth}
		\centering
		\begin{minipage}[c]{.025\textwidth}
			\subcaption{}
			\label{fig:impl_c}
		\end{minipage}
		~%
		\begin{minipage}[c]{.65\textwidth}
			\begin{tikzpicture}
				\tikzstyle{every node}=[
				node font=\sffamily\footnotesize,
				minimum height=\nodeminheight,
				minimum width=\nodeminwidth,
				inner sep=2mm,
				align=center,
				node distance=\nodedistvert and \nodedisthor,
				]
				\tikzstyle{comp}=[
				fill=white,
				drop shadow,
				draw,
				]
				\def\size{4*\nodeminwidth+3*\nodedisthor}
				\def\shiftdist{.25cm}
				\def\appresNodeShift{-20mm+3*\shiftdist}
				\node[comp] (tom) [anchor=base,
					general shadow={
							shadow scale=1,
							shadow xshift=1ex,
							shadow yshift=1ex,
							drop shadow={
									shadow xshift=1.5ex,
									shadow yshift=.5ex,
								},
							drop shadow={
									shadow xshift=1ex,
									shadow yshift=0ex,
								},
							draw,
							fill=white,
						},
					general shadow={
							shadow scale=1,
							shadow xshift=.5ex,
							shadow yshift=.5ex,
							draw,
							fill=white,
						}
				] {\guillemotleft\textit{PEP}\guillemotright\\TOM};
				\node[comp] (app) [left=of tom] {Untrusted\\Application Logic};
				\node[comp] (tps) [right=of tom] {\guillemotleft\textit{PDP}\guillemotright\\TPS};
				\node[comp] (tep) [right=of tps] {\guillemotleft\textit{EPP}\guillemotright\\TEP};
				\draw[
					->,
					>=latex,
					semithick,
				]
				([yshift=+\shiftdist]app.east) edge ([yshift=+\shiftdist]tom.west)
				([yshift=-\shiftdist]tom.west) edge ([yshift=-\shiftdist]app.east)
				([yshift=+\shiftdist]tom.east) edge ([yshift=+\shiftdist]tps.west)
				([yshift=-\shiftdist]tps.west) edge ([yshift=-\shiftdist]tom.east)
				([yshift=+\shiftdist]tps.east) edge ([yshift=+\shiftdist]tep.west)
				([yshift=-\shiftdist]tep.west) -- ([yshift=-\shiftdist]tps.east)
				;
				\begin{scope}[on background layer]
					\draw[thick,densely dotted,fill=\boxfillcolor]($(app.north west)+(-\boxboarder,\boxboarder)$) rectangle ($(app.south east)+(\boxboarder,-\boxboarder)$);
					\draw[thick,densely dotted,fill=\boxfillcolor] ($(tom.north west)+(-\boxboarder,\boxboarder)$) rectangle ($(tep.south east)+(\boxboarder,-\boxboarder)$);
				\end{scope}
			\end{tikzpicture}
			\vspace{\spacebetweenvariants}
		\end{minipage}
	\end{subfigure}
	\begin{subfigure}[!tb]{.85\textwidth}
		\centering
		\begin{minipage}[c]{.025\textwidth}
			\vspace{6pt}%
			\subcaption{}
			\label{fig:impl_d}
		\end{minipage}
		~%
		\begin{minipage}[c]{.65\textwidth}
			\begin{tikzpicture}
				\tikzstyle{every node}=[
				node font=\sffamily\footnotesize,
				minimum height=\nodeminheight,
				minimum width=\nodeminwidth,
				inner sep=2mm,
				align=center,
				node distance=\nodedistvert and \nodedisthor,
				]
				\tikzstyle{comp}=[
				fill=white,
				drop shadow,
				draw,
				]
				\def\size{4*\nodeminwidth+3*\nodedisthor}
				\def\shiftdist{.25cm}
				\def\appresNodeShift{-20mm+3*\shiftdist}
				\node[comp] (tom) [anchor=base,
					general shadow={
							shadow scale=1,
							shadow xshift=1ex,
							shadow yshift=1ex,
							drop shadow={
									shadow xshift=1.5ex,
									shadow yshift=.5ex,
								},
							drop shadow={
									shadow xshift=1ex,
									shadow yshift=0ex,
								},
							draw,
							fill=white,
						},
					general shadow={
							shadow scale=1,
							shadow xshift=.5ex,
							shadow yshift=.5ex,
							draw,
							fill=white,
						}
				] {\guillemotleft\textit{PEP}\guillemotright\\TOM};
				\node[comp] (app) [left=of tom] {Untrusted\\Application Logic};
				\node[comp] (tps) [right=of tom] {\guillemotleft\textit{PDP}\guillemotright\\TPS};
				\node[comp] (tep) [right=of tps] {\guillemotleft\textit{EPP}\guillemotright\\TEP};
				\draw[
					->,
					>=latex,
					semithick,
				]
				([yshift=+\shiftdist]app.east) edge ([yshift=+\shiftdist]tom.west)
				([yshift=-\shiftdist]tom.west) edge ([yshift=-\shiftdist]app.east)
				([yshift=+\shiftdist]tom.east) edge ([yshift=+\shiftdist]tps.west)
				([yshift=-\shiftdist]tps.west) edge ([yshift=-\shiftdist]tom.east)
				([yshift=+\shiftdist]tps.east) edge ([yshift=+\shiftdist]tep.west)
				([yshift=-\shiftdist]tep.west) -- ([yshift=-\shiftdist]tps.east)
				;
				\begin{scope}[on background layer]
					\draw[thick,densely dotted,fill=\boxfillcolor]($(app.north west)+(-\boxboarder,\boxboarder)$) rectangle ($(app.south east)+(\boxboarder,-\boxboarder)$);
					\draw[thick,densely dotted,fill=\boxfillcolor] ($(tom.north west)+(-\boxboarder,\boxboarder)$) rectangle ($(tom.south east)+(\boxboarder,-\boxboarder)$);
					\draw[thick,densely dotted,fill=\boxfillcolor] ($(tps.north west)+(-\boxboarder,\boxboarder)$) rectangle ($(tep.south east)+(\boxboarder,-\boxboarder)$);
				\end{scope}
			\end{tikzpicture}
		\end{minipage}
	\end{subfigure}
	\caption{\AppSPEAR{} instantiations with different separation/isolation of the architecture components.}
	\label{fig:impl}
	\Description[<short description>]{<long description>}
\end{figure*}

\section{Architecture Implementation}

\AppSPEAR{} provides a functional framework for implementing the reference monitor principles at application layer. The flexibility of \AppSPEAR{} enables the separation of its architecture components and their implementation by means of isolation mechanisms in a variety of ways. §\,\ref{sec:archimpl:balancing} discusses multiple architecture instantiation alternatives and their individual characteristics regarding TCB size. Each variant has, on the one hand, a certain degree of possible rigor in terms of TCB isolation (\textit{effectiveness}) and, on the other hand, certain costs in terms of required resources and communication effort for crossing isolation boundaries (\textit{efficiency}). To implement these instantiation alternatives, a selection of isolation mechanisms and developer support provided by a configurable developer framework are discussed. §\,\ref{sec:archimpl:study} then describes a practical case study integrating \AppSPEAR{} into the electronic medical record system OpenMRS.

\subsection{Balancing Effectiveness and Efficiency}
\label{sec:archimpl:balancing}

\paragraph*{Architecture Instantiation.}
\AppSPEAR{}'s design provides different alternatives for instantiating and implementing its components. Fig.~\ref{fig:impl} shows reasonable variants illustrating boundaries at which the architecture components are isolated from one another (by isolation mechanisms). Since the TEP realizes parts of the implementation of context-based security policies (\eg{} risk analysis and estimation), TPS and TEP are explicitly not isolated from each other.

If a large application TCB comprising the potentially untrusted application logic as well as all \AppSPEAR{} components (see Fig.~\ref{fig:impl_a}), then the implementation is conceptually and, thus, both in terms of strictness and costs at the level of application"=integrated security policies. From a qualitative point of view, \AppSPEAR{} enables a structured software engineering of applications to be equipped with individual security policies.

Beyond this simple low"=cost but low"=quality implementation, the security architecture supports several alternatives with greater effectiveness. Taking advantage of the functional encapsulation of the TPS and the TEP, the security policy and its runtime environment can be implemented isolated from the rest of the application (see Fig.~\ref{fig:impl_b}). This paves the way for a tamperproof~-- naturally depending on the strength of the used isolation mechanism~-- and analyzable policy implementation. In quantitative terms, compared with the previous alternative, the application TCB comprises still the same set of functions, but here, the integrity of the security policy is protected by a more effective isolation.

A further qualitative improvement of the effectiveness is achieved by also isolating the TOMs as part of the application TCB from the untrusted application (see Fig.~\ref{fig:impl_c}). The isolation boundary between the TOMs and the TPS/""TEP is moved between the untrusted application logic and the TOMs. This results in a smaller application TCB perimeter due to the fact that only security"=relevant functions are part of the TCB. From a conceptual perspective, the costs are comparable to the second variant (see Fig.~\ref{fig:impl_b}) due a single isolation boarder and an equally frequent crossing of the TCB boundary. Nevertheless, the software engineering effort that is required for existing software architectures to create the technical prerequisites, \ie{} a well-defined interface between untrusted application logic and TOMs, is also relevant when considering the isolation of TOMs. Since many applications typically use a database system and access stored object data, there is often already a functionally separated object management as intended by TOMs, which puts the costs into perspective at this point.

The example of an application utilizing a database already shows that due to the size and complexity of today's application TCBs and, consequently, the high verification effort, not the entire TCB can always be proven to be correctly implemented using conventional methods. Analogous to the principle of separate server processes in microkernel"=based OSs, the TPS can be isolated as an essential part of the TCB from the rest of the TCB (see Fig.~\ref{fig:impl_d}) for security and robustness reasons (\eg{} preventing unintended changes to the policy data structures due to reference errors). Thus, correctness properties of the security policy can be analyzed easier by formal models and methods (\eg{} security properties such as right proliferation (safety) in dynamic AC models). This qualitative improvement also implies additional costs, since isolation mechanisms are used at two borders and these have to be overcome in a controlled manner for each policy request.

\paragraph*{Isolation Mechanisms.}
In order to get beyond the weak security guarantees of application-integrated security policies (cf. Fig.~\ref{fig:impl_a}), harder measures are necessary. Based on the classification of isolation mechanisms in \cite{SWG+16}, we select and discuss the implementation of \AppSPEAR{} (cf. Fig.~\ref{fig:impl_b}\,--\,\ref{fig:impl_d}) by means of two mechanisms that go beyond purely software-based, intra-application-enforced isolation.

At the OS level, the process abstraction together with the virtual memory management provides one of the most fundamental isolation mechanisms. Each process has its own private virtual address space so that the memory of executed programs is isolated. This allows not only to allocate memory resources as optimally as possible according to processes' needs, but also to avoid the propagation of errors, faults, and failures to other processes or the entire system. Futhermore, even if a process is compromised, the adversary cannot breach the security of other processes without extensive efforts. By isolating the \AppSPEAR{} components from the potentially untrusted application logic and from each other using virtual address spaces via processes as in Fig.~\ref{fig:impl_b}\,--\,\ref{fig:impl_d}, bugs and security vulnerabilities of the untrusted application part can no longer simply affect the trusted parts. To enable communication across process boundaries, communication mechanisms controlled by the OS are necessary, such as local domain sockets, messages queues, pipes etc., which may be chosen according to the criteria of performance, flexibility, and usability.

Beyond that, Intel SGX enables applications to protect private code and data from privileged system software such as the OS kernel, hypervisor, and BIOS as well as other applications. To achieve this, SGX uses protected TEEs called enclaves. An enclave is a protected area within an application process' address space. To meet integrity and confidentiality requirements, protected application code is loaded into an enclave after measuring using hardware-based attestation, and enclave data is automatically encrypted when leaving the CPU package into memory. Consequently, this design significantly reduces the TCB to only code executed inside the enclave (as well as the CPU, which in the end must always be trusted).

Since enclave memory cannot be read directly from outside of an enclave, data, which needs to be passed between trusted and untrusted application parts, has to be copied explicitly from and to an enclave. The SGX SDK provides mechanisms to create corresponding bridge functions, \emph{ecalls}, which dispatch enclave entry calls to corresponding functions inside the enclave defining an enclave's interface. Corresponding functions that reside in the untrusted application part are called \emph{ocalls} and invoked inside the enclave to request services outside of the enclave (\eg{} system calls).

\paragraph*{Software Framework and Developer Support.}
The adaption and implementation of \AppSPEAR{} is supported by a software developer framework. As discussed, \mbox{hardware-} and encryption"=based trusted execution mechanisms provide a promising basis for a trusted enforcement of application security policies. Specifically, we include support for Intel SGX due to its popularity and availability in current Intel CPUs. Since the official SGX SDK is only available in C++, we implemented the framework in C++ as well. Our \AppSPEAR{} developer framework comprises the following features:
\begin{enumerate}[label=\theenumi.]
	\item transparent and flexibly configurable isolation according to the \AppSPEAR{} instantiation variants and the selection of isolation mechanisms discussed,\footnote{Note that the framework implementation is conceptually not limited to this selection and can be easily extended to support additional isolation mechanisms.}
	\item transparent and flexibly configurable RPC/""RMI-alike communication
	      via proxy objects,
	\item transparent reduction of the communication effort via
	      in-proxy caching.
\end{enumerate}

The key idea for implementing these features lies in an RPC/""RMI-""alike communication model:
Any communication between the components of a policy-controlled application is handled via pairs of proxy objects, similar to stubs known from RPC/""RMI implementations. Those proxys act as intermediaries representing the callee on the caller side and the caller on the callee side. Consequently, any communication handled by these proxys is be performed \emph{transparently}, so that the consideration of isolation-""mechanism-""specific needs is not required to be considered by developers, such as the serialization of application-specific data structures and accordingly their deserialization. In addition, techniques for securing communication, such as encryption, hashing, or integrity certificates, and reducing communication effort, such as caching of access requests and corresponding decisions, are transparently integrated into proxys.

The developer framework provides a set of requester/responder proxy pair implementations, one for each \AppSPEAR{} component taking the specific characteristics of each isolation and corresponding communication mechanism into account:
\begin{itemize}
	\item language/compiler-based isolation and communication via local procedure/function/method calls,
	\item process-based isolation and IPC-based communication via local domain sockets\footnote{Sockets were chosen based on the out-of-the box compromise of flexibility (in terms of message content/size), performance, simplicity, and availability in commodity OSs.}, and
	\item TEE-based isolation via SGX enclaves and cross-""enclave-""boundary communication (ecalls/""ocalls).
\end{itemize}
Base classes for all parts of a security-policy-controlled application provide basic functionality as explained in §\,\ref{sec:archdesign:archdesign} and abstract member objects for proxy-based communication. Concrete implementations are derived from those base classes where their instantiation also configures the particular isolation and communication mechanisms to be used, which is usually done at compile time. Alternatively, to be able to compare different implementation variants (\eg{} for testing purposes), the configuration can be done flexibly at runtime (see also §\,\ref{sec:eval}).

In order to reduce communication effort and multiple processing of the same inputs, caching mechanisms can be used. In principle, caching is possible for
\begin{ilenum}
	\item application-logic-to-TOM communication, and
	\item TOM-to-TPS communication.\footnote{Communication between the TPS and the TEP is asynchronous based on callbacks or periodic updates (see §\,\ref{sec:archdesign:archdesign}). Since received context values are temporarily stored anyway, a separate cache is not useful here.}
\end{ilenum}
Since the application logic is assumed to be untrusted, caching of TOM calls and results on the application logic side counteracts our goals of a rigorous and trustworthy policy enforcement. Nevertheless as argued in §\,\ref{sec:archimpl:balancing}, many applications use database systems to store application objects such that database management system (DBMS) caches can achieve an overhead reduction when retrieving application object data.

Caching of TPS requests and decisions on the TOM side achieves to avoid multiple identical requests (if the policy has not changed compared to the initial request) and reduces the runtime overhead to the level of local function/method calls~-- especially for implementation variants with stricter TPS/TEP isolation (see Fig.~\ref{fig:impl_b} and \ref{fig:impl_d}). In order to also keep access times to cache entries as low as possible, the cache is implemented using a container organized on the basis of a hash table. If, in the case of a dynamic security policy, the execution of a state-modifying operation results in the modification of a component (\eg{} assignment or activation of an additional role in a dynamic RBAC model), it may be necessary to invalidate one or more cache entries, the TPS initiates an invalidation of the respective cache entry.

\subsection{Case Study: Electronic Medical Record System OpenMRS}
\label{sec:archimpl:study}

This section describes a case study in which we apply the \AppSPEAR{} framework to an existing database-""backed application. Subject of the study is OpenMRS, an open source electronic medical record (EMR) and medical information management system \cite{OpenMRSa}. OpenMRS is an optimal representative for a policy-controlled application, since the AC policy is part of its architecture and directly visible in the source code.

\paragraph*{Software and Security Architecture.}
OpenMRS aims at being adaptable to resource"=constrained environments such as healthcare facilities of low-income countries \cite{WMB+06,TAB+10,TCL+10}. This motivation is reflected in the modular web application architecture which consists of three logical layers:
\begin{enumerate}[label=\theenumi.]
	\item the \emph{user interface (UI) layer}, providing the UI via input and query forms,
	\item the \emph{service layer}, implementing the basic functionality (including data model interaction) and a corresponding API, and
	\item the \emph{database layer}, realizing the anchored data model.
\end{enumerate}

A tightly integrated role-""based AC (RBAC) policy is responsible for controlling accesses to application objects such as EMRs or medication plans. The policy semantics are similar to the RBAC model family \cite{Ferraiolo07a}: a logged-in user's accesses to objects are regulated according to her assigned and activated roles (\eg{} physician or nurse), to which certain permissions for actions are assigned (\eg{} \textit{read}/\textit{modify} \textit{EMR} or \textit{create}/\textit{delete} \textit{patient}).

RBAC policy and enforcement are implemented through an AOP mechanism (\texttt{AuthorizationAdvice}) that wraps each service layer method call with a policy call, and custom Java annotations (\enquote{\texttt{@Authorized}}) which initiate checking the privileges of the currently authenticated user. Since required permissions are directly attached to each service layer method, the policy is hard-coded and distributed over 635~points (OpenMRS core module version~2.3.1). This renders the policy as well as its underlying model static contradicting the goal of simple adaptability and flexible configurability. Beyond that, policy decision and enforcement functionality is isolated from application logic only via language-based mechanisms and, thus, misses the opportunity of using stronger isolation for the correct and manipulation-free handling of critical medical data.

\paragraph*{Applying the \AppSPEAR{} Framework.}
While OpenMRS is implemented in Java, the official SGX SDK and the \AppSPEAR{} developer framework are implemented in C++. Therefore, for the scope of this study, we decided to prototype OpenMRS in C++.

While maintaining the software architecture, a few functional differences exist. On the UI layer, we have implemented a minimal command line interface which is sufficient for requesting service layer functionality. On the service layer, we have implemented a selection of core services for data model interaction: the (system) user service (\textit{login}, \textit{activate/deactivate role}, \textit{logout}), the person service (\textit{create/delete person object}, \textit{get/set address}), and the patient (EMR) service (\textit{create/delete patient object}, \textit{get/set patient diagnosis}), each derived in an OOP manner from an abstract service class. On the database layer, SQLite \cite{SQLite19a} is used instead of MySQL as relational DBMS because of its ability to fully store a database in-memory enabling its trusted execution within an SGX enclave. We modified SQLite version 3.32.3 for an in-enclave execution by wrapping system calls through trampoline functions that temporarily exit an enclave (\emph{ocalls}) or, where possible, by an SGX-compatible variant provided by the SDK.

Each service forms a TOM and manages its own objects (users, patients, EMRs, etc.) stored in the database and provides corresponding operations on them (generally such as \textit{create/""destroy object} and \textit{read/""modify object attribute}). When using the communication proxys provided by our developer framework, operations on TOM-managed objects called within the UI layer are transparently forwarded to the proxy counterpart of the respective TOM; depending on the isolation mechanism used, \eg{} either via function/method calls, local domain socket send/""receive (IPC) or SGX enclave calls). An analogous pattern applied for the TOM-""to-""TPS communication initiates requests to the RBAC policy regarding access permission or denial for each TOM operation to be executed. The TPS can realize the security policy by means of either a database or, leading to a smaller TCB, model-tailored data structures (as in our implementation).


\section{Evaluation}
\label{sec:eval}

This section presents an evaluation of \AppSPEAR{} based on its reference implementation. The evaluation addresses the practical feasibility of the application-level approach in terms of the runtime performance impact of the different instantiation alternatives. Each alternative is implemented by configuring our developer framework to use each of the following isolation mechanisms:
\begin{ilenum}
	\item language/compiler-based isolation serving as a baseline (intra-application) mechanism,
	\item process-based isolation as a basic OS-level mechanism, and
	\item TEE/enclave-based isolation by use of Intel SGX as a widely available hardware-level trusted execution technology.
\end{ilenum}
In particular, we study two use cases: A synthetic application serves as a baseline for the costs of applying \AppSPEAR{} whereas our prototypical reimplementation of OpenMRS (see §\,\ref{sec:archimpl:study}) illustrates the costs in terms of runtime overhead in a practical scenario. Subsequently, §\,\ref{sec:eval:meth} describes the evaluation method in detail and §\,\ref{sec:eval:res} discusses the results.

\subsection{Evaluation Methodology}
\label{sec:eval:meth}

The practical feasibility is examined by measuring the runtime performance of each architecture variant and the mentioned isolation mechanisms.

\paragraph*{Test Cases}
A simple application with a synthetic always-""allow security policy highlights the baseline for the runtime which has to be considered for isolation and communication in each architecture implementation variant (baseline microbenchmark). The application only includes a single synthetic operation passing an operation identifier (\eg{} when using SGX-based isolation, a minimum amount of copying is required to determine the lowest cost for entering an enclave).

Our prototypical reimplementation of OpenMRS in C++ with an RBAC policy (see §\,\ref{sec:archimpl:study} for details) serves as a real-world use case. In particular, the layered software architecture and the usage of a database is representative for a multitude of applications, so that this use case also yields a good impression of potential costs also relevant for other scenarios. We use a database provided by the OpenMRS community comprising an anonymized data set of 5,000 patients and 500,000 observations~\cite{OpenMRSc}.

We run two types of benchmarks: four microbenchmarks show the effort for typical create, read, update, and destroy (CRUD) operations of selected OpenMRS services, whereas a mixed macrobenchmark based on an OpenMRS workload extracted from logs \cite{CSH+19} (called \enquote{Action} there) puts the execution of individual operations into a bigger context. Our adapted workload assumes the following proportional occurrences of patient service CRUD operations:
\begin{itemize}
	\item 25\,\% of create patient (object) operations (additionally create person is called on person service if a corresponding person object does not yet exist which is the case here),
	\item 38\,\% of read patient diagnosis operations,
	\item 12\,\% of update patient diagnosis operations, and
	\item 25\,\% of delete patient (object) operations.
\end{itemize}
The individual operations appear mixed over the entire execution.
To make the runtimes comparable to those of individual operations, the measured runtimes are divided by 100 (number of executed operations within our macrobenchmark).

Since research has shown that, in particular, TEE/enclave-based isolation is expected to involve high costs \cite{WAK18}, we have also implemented and evaluated two techniques for decreasing runtime overhead:
\begin{ilenum}
	\item transparent caching of policy decisions in TPS proxies located in the TOMs (as discussed in §\,\ref{sec:archimpl:balancing}) and
	\item asynchronous enclave calls \cite{ATG+16,OLM+17,WBA17} which are provided by the SGX SDK as \emph{switchless calls} because they do not involve costly enclave switches.
\end{ilenum}

\begin{figure*}[b!]
	\centering
	\begin{subfigure}[!tb]{.485\textwidth}
		\includegraphics[width=\textwidth,trim=7pt 8pt 30pt 26pt,clip]{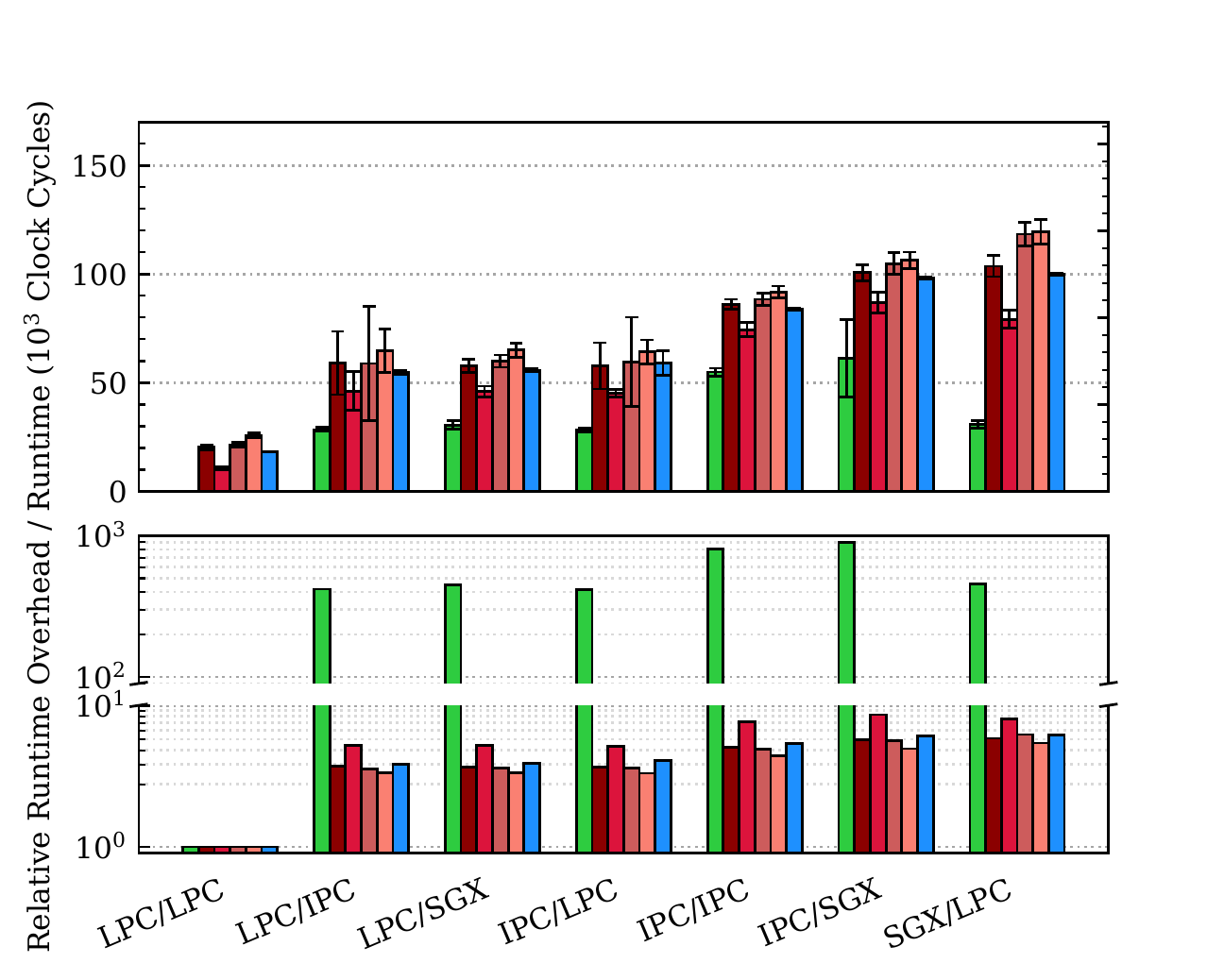}
		\subcaption{}
		\label{fig:plot_bench_a}
	\end{subfigure}
	~
	\begin{subfigure}[!tb]{.485\textwidth}
		\includegraphics[width=\textwidth,trim=30pt 8pt 7pt 26pt,clip]{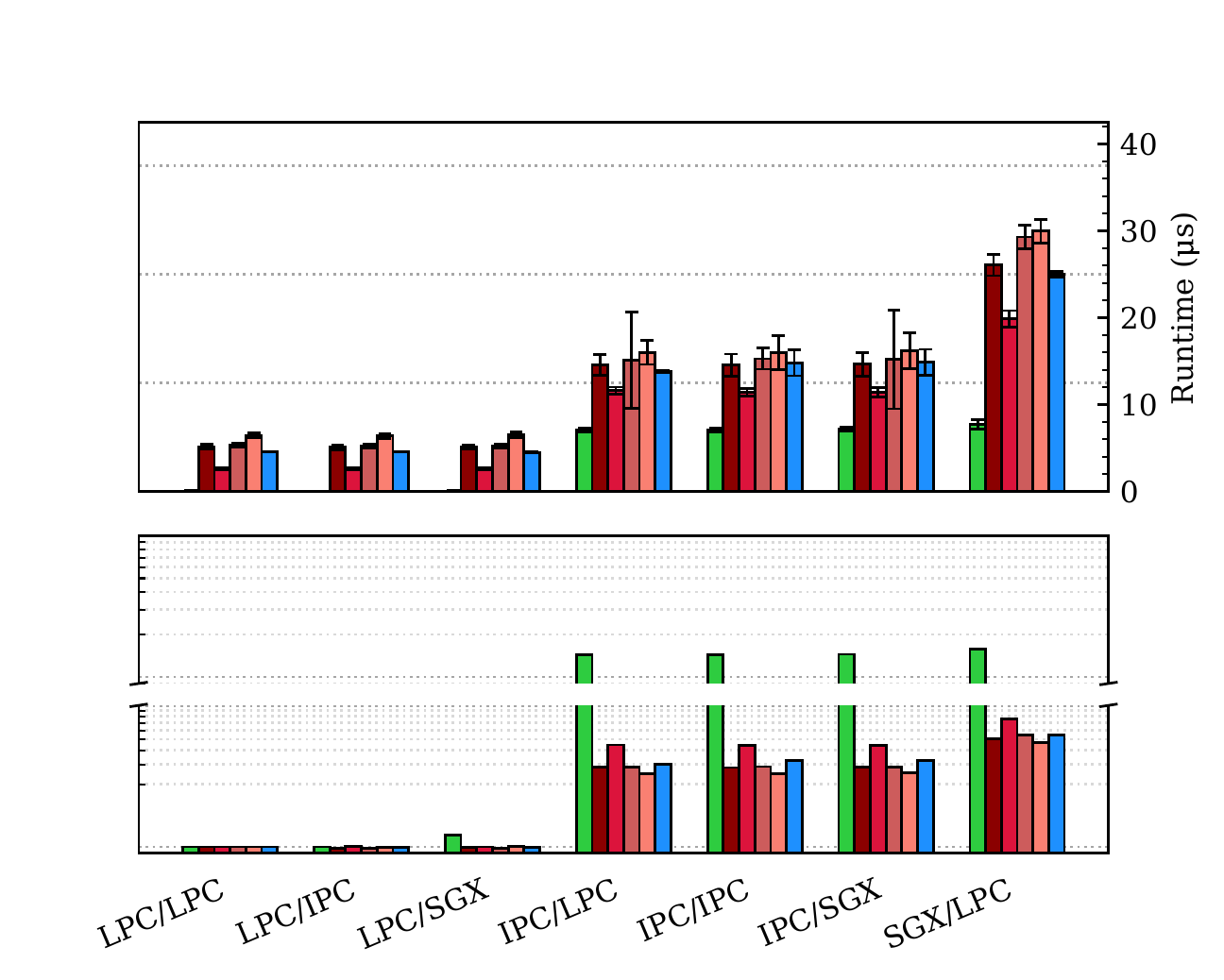}
		\subcaption{}
		\label{fig:plot_bench_b}
	\end{subfigure}
	\vspace{.01\baselineskip}

	\begin{subfigure}[!tb]{.485\textwidth}
		\includegraphics[width=\textwidth,trim=7pt 8pt 30pt 26pt,clip]{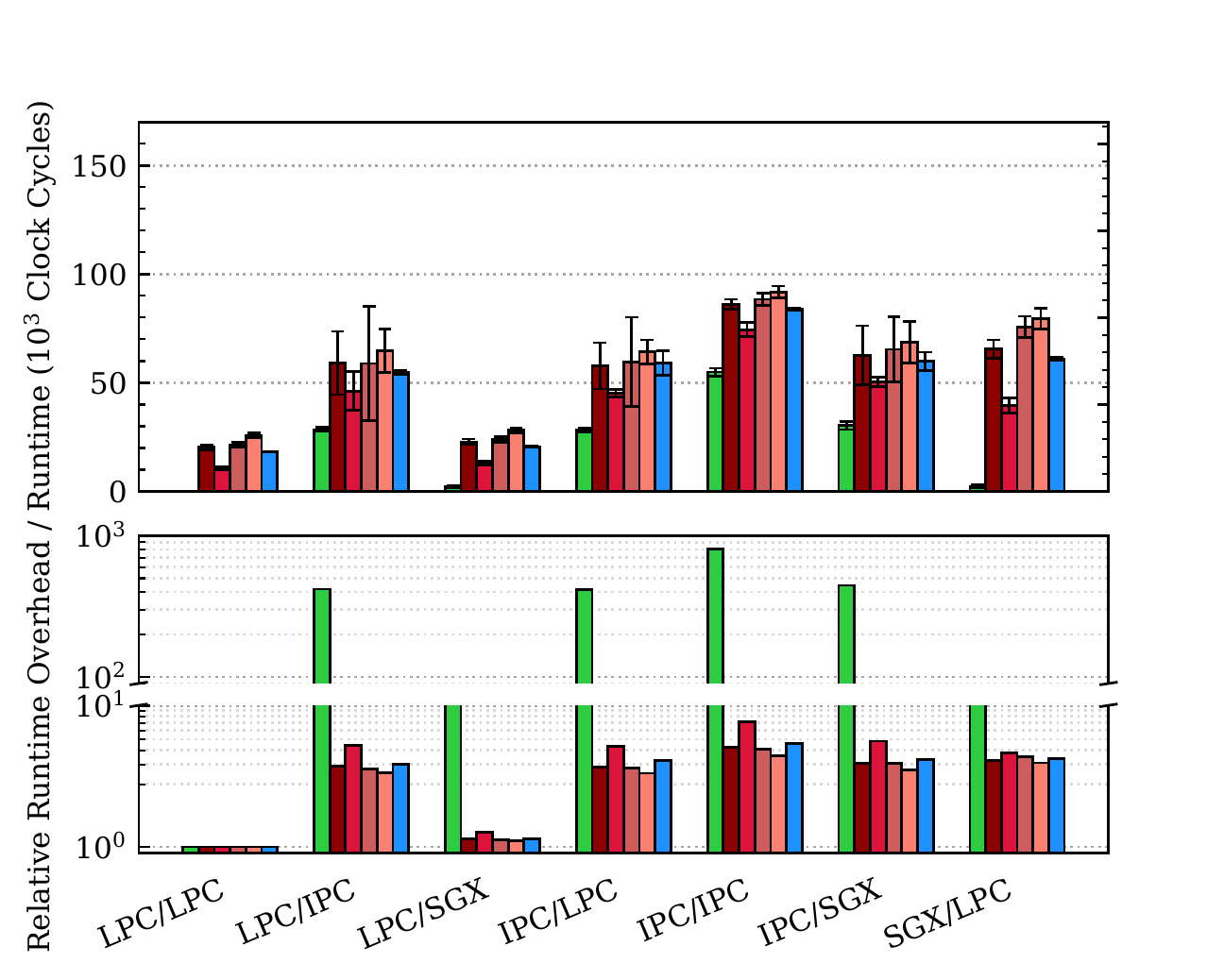}
		\subcaption{}
		\label{fig:plot_bench_c}
	\end{subfigure}
	~
	\begin{subfigure}[!tb]{.485\textwidth}
		\includegraphics[width=\textwidth,trim=30pt 8pt 7pt 26pt,clip]{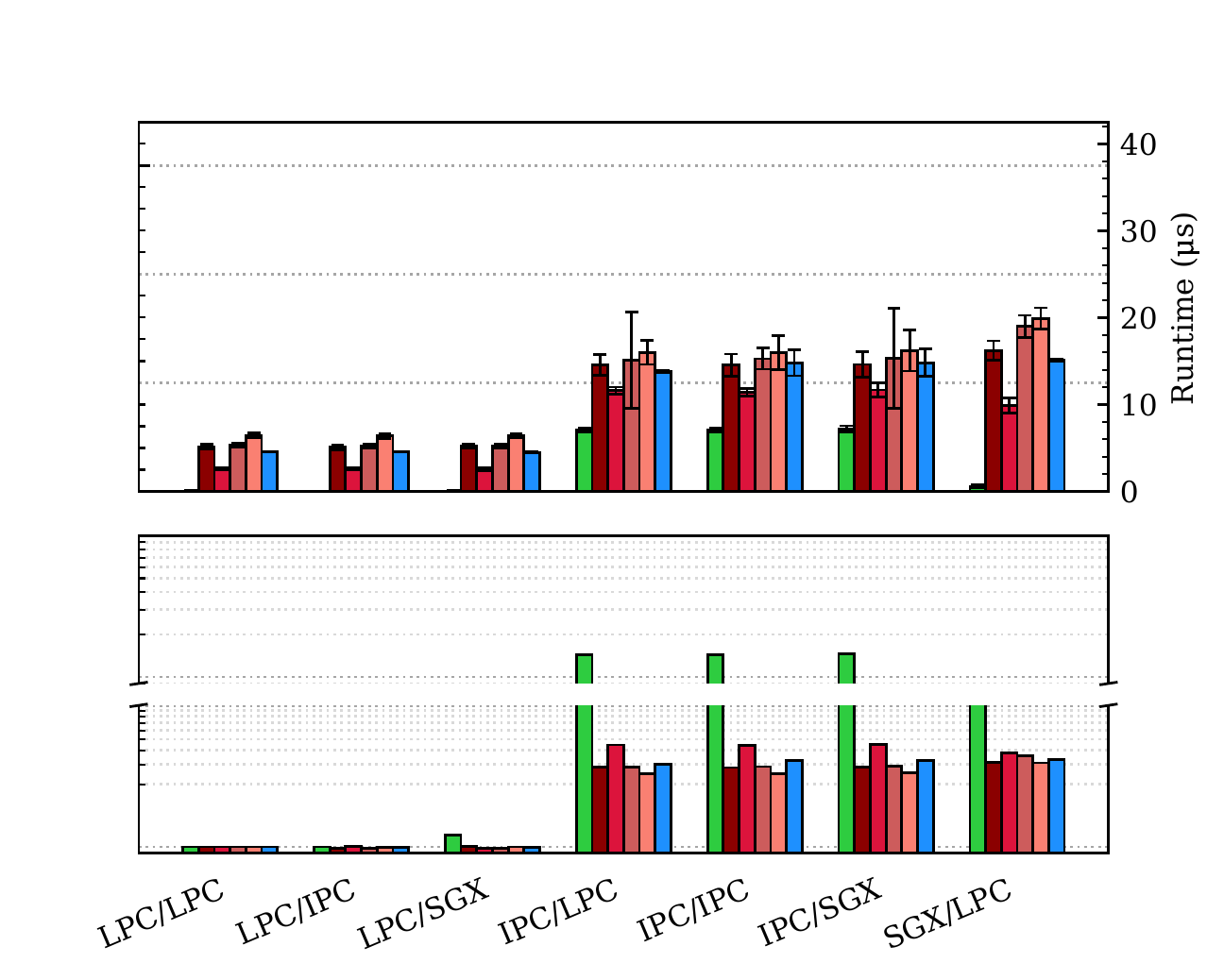}
		\subcaption{}
		\label{fig:plot_bench_d}
	\end{subfigure}
	\vspace{.7\baselineskip}

	\vspace{2pt}
	\includegraphics[trim=1pt 181pt 1pt 1pt,clip]{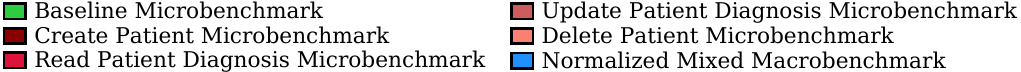}

	\caption{Benchmarking results: (a)~standard/unoptimized implementations, (b)~implementations optimized by caching, (c)~implementations optimized by switchless calls, (d)~implementations optimized by caching and switchless calls.}
	\label{fig:plot_bench}
	\Description[<short description>]{<long description>}
\end{figure*}

\paragraph*{Metrics and Measurements}
The runtimes are measured in CPU clock cycles by using the RDTSCP instruction. To avoid typical behavior of \enquote{cold} CPU caches, we perform the measurements only after a warm-up phase of 1,000,000 iterations. To filter out potential outliers which typically occur, we perform 1,000,000 iterations for each measurement and calculate the median.

All measurements were performed on desktop hardware with an Intel Core i7-7700K CPU at 4.2 GHz and 32~GiB DDR4 RAM at 2,400 MHz. The machine runs Ubuntu 18.04.4 LTS (64-bit) with Linux kernel version 5.3.0-59 including mitigations (kernel and microcode updates) for Meltdown, Spectre, L1TF/Foreshadow, and MDS class vulnerabilities \cite{KHF+19,LSG+18,BMW+18,SMO+19,CGG+19,SLM+19} which have shown a runtime of over $2.2$ times more than without patches \cite{WAK18}. We use the Intel SGX driver, SDK and platform software (PSW) in version 2.7.1 \cite{Int19a}. We compile using GCC 9.2.0, SGX hardware mode, and the SGX SDK Prerelease configuration known to have the same performance as production enclaves \cite{Int16a}. The Enclave Page Cache (EPC) size is set to the maximum of 128 MiB of which approximately 93 MiB are usable. During all experiments we disable dynamic CPU frequency scaling, Turbo Boost, and Hyper-Threading to avoid erratic runtime behavior. To further reduce potential outliers, we adjust both process scheduling priorities and interrupt affinities.

\subsection{Evaluation Results}
\label{sec:eval:res}

This section discusses the results of the experimental evaluation. For each \AppSPEAR{} implementation variant (labels on the x-axis are chosen according the type of isolation/communication used between application logic and TOMs as well as between TOMs and TPS/TEP), Figs.~\ref{fig:plot_bench_a}\,--\,\ref{fig:plot_bench_d} show absolute runtimes in the upper part (unit ${10}^3$ clock cycles on the left y-axis, unit microseconds calculated for the used Intel i7-7700K CPU with a base frequency of 4.2 GHz on the right y-axis) and in the lower part the relative runtime overhead compared to the fully integrated, intra-application implementation of \AppSPEAR{}. The error bars show 95\,\% confidence intervals.

First of all, Fig.~\ref{fig:plot_bench_a} illustrates the results of the baseline microbenchmarks showing the basic runtime overhead required for isolation and communication in each of the considered unoptimized \AppSPEAR{} implementations. Each implementation variant requires at least a relative runtime overhead of more than 2 orders of magnitude compared to the intra-application implementation of \AppSPEAR{}. Due to isolation and communication effort occuring twice in the IPC/IPC and IPC/SGX variants, the costs are also about twice as high compared to the single process- and SGX-isolated variants (ca. 30k cycles vs. ca. 55k/60k cycles).

Considering the CRUD microbenchmarks, in the LPC/LPC implementation, the read operation requires only about half of the runtime of create, update and destroy operations (ca. 10,5k cycles vs. ca. 18,2k/""21,5k/""25,8k cycles) because no changes to application objects stored in the database are made. Using process-based isolation in the LPC/IPC, IPC/LPC, and IPC/""IPC implementation variants, a comparable picture emerges when the costs of isolation and communication are taken into account. The LPC/IPC and IPC/LPC variants yield a similar level of runtime overhead, while the IPC/IPC variant has an increased amount of additional overhead due to isolation using two separate processes which nevertheless does not double the costs due to a compact small-volume TOM-to-""TPS communication.

While for the same reason, the SGX-based isolation of the TPS/""TEP (LPC/SGX variant) has a runtime overhead comparable to the LPC/IPC variant, the SGX/LPC implementation variant highlights that the joint isolation of TOM and TPS/TEP and the resulting in-enclave code execution can lead to a considerable additional overhead beyond pure isolation/""communication costs: The approximately twice as high runtime effort results from the execution of SQLite within the isolated enclave, since temporary enclave exits (ocalls) may occur several times (see also §\,\ref{sec:archimpl:study}).

The hybrid IPC/SGX variant shows a middle way solution in rigor and costs compared to the previously discussed variants: TOMs are separated from the application logic by process-based isolation and only the TPS/TEP is completely trusted through isolation via SGX. Although the baseline overhead is the highest in the field of comparison, the CRUD microbenchmark results rank between the IPC/IPC and SGX/LPC variants because SQLite is executed in a regular process. Moreover, the macrobenchmark results fit completely into the picture of results drawn so far without any anomalies. The runtimes for the CUD operations in favor of the R operation are relativized in all \AppSPEAR{} implementation variants according to a real-world workload.

In order to reduce the runtime overhead (especially the communication overhead), we have implemented and evaluated two techniques for runtime reduction, whose impact we will now discuss: caching of policy requests and decisions within TOMs (see Fig.~\ref{fig:plot_bench_b}) as well as SGX switchless calls (see Fig.~\ref{fig:plot_bench_c}). The runtimes of cache-optimized \AppSPEAR{} implementations assume caches already filled with corresponding entries (done in measurement warm-up).

It can be seen in Fig.~\ref{fig:plot_bench_b} that especially those implementation variants that separate the TPS/TEP from TOMs via process- or SGX-based isolation (LPC/IPC, LPC/SGX, IPC/IPC and IPC/SGX implementation variants) benefit from this measure: A cache hit effectively results in a significant reduction of the runtime overhead by eliminating IPC/socket or SGX/enclave communication. For the LPC/IPC and LPC/SGX variants, the runtime is reduced to the level of the intra-application implementation (LPC/LPC variant) and for the IPC/SGX and IPC/IPC variants to the cost level of IPC/LPC. For all other implementation variants (IPC/LPC, IPC/IPC, and SGX/LPC) querying the cache requires a little more runtime.

Compared to the \AppSPEAR{} standard implementations, SGX switchless calls improve runtimes in all considered implementation variants which use SGX/enclave-based isolation/communication, \ie{} LPC/""SGX, IPC/""SGX, and SGX/""LPC (see Fig.~\ref{fig:plot_bench_c}). For the LPC/SGX and SGX/LPC variants, this approach leads to a reduction of the basic runtime overhead by more than one order of magnitude, while for the IPC/""SGX variant it is more than half of the runtime. The results of the CRUD micro- and macrobenchmarks also benefit from this: The runtimes are reduced on average by more than half for the LPC/SGX variant and by less than two thirds for the other two IPC/SGX and SGX/LPC variants. The results of all other variants remain unchanged compared to Fig.~\ref{fig:plot_bench_a}.

Finally, Fig.~\ref{fig:plot_bench_d} shows the results of combining the two previously discussed optimizations. Taking into account both caching (assuming cache hits) and switchless calls, the runtimes for TPS/TEP isolating implementation variants via processes or SGX (LPC/IPC and LPC/SGX) are reduced to the cost level of an intra-application implementation (LPC/LPC variant). Although the other implementation variants have 3 to 4 times higher runtimes, due to a much higher effort for communication and isolation, they also allow for much stronger isolation guarantees (see also §\,3.1). In particular, the trusted execution of all \AppSPEAR{} components within an enclave (SGX/LPC variant) is now possible almost at the run-time level of process-based isolation, putting the expected costs of using Intel SGX trusted execution technology for our application-level security policy enforcement approach into perspective.


\section{Related Work}

This section summarized works related to our application-level policy enforcement approach considering a precise TCB perimeter.

A step towards more precisely identifiable TCB perimeters of applications is observable in SELinux which provides user-space object managers \cite{LS01a}. The approach is based on auxiliary application"=level constructs for managing application objects as SELinux policy objects and for supplementing the system-wide security policy with application policy rules based on OS-specific policy semantics \cite{Car07,Wal07,Koh11}. Although collecting and locating all policy rules in the kernel's policy runtime environment is more effective in terms of policy protection and analyzability, this approach causes an increase of each application TCB by the OS policy as well as its runtime functionality and, above all, gives up any policy individuality.

An approach that goes beyond this is the SELinux Policy Server Architecture \cite{Tresys,MBM+06}. With the goal of a clear separation between OS-level and application-level policies, the user-space security server (USSS) is placed at application level running all application-specific policies. Although the USSS fulfills a comparable role to the \AppSPEAR{} TPS, the USSS exists only once at application level and is not instantiated individually per application (in contrast to the TPS). Consequently, this approach leads to considerably larger application TCBs and, additionally, has to deal with the problems known from multi- and meta-policy systems \cite{Bonatti02a}. Due to unknown reasons the project was abandoned.


\section{Conclusion}

This paper tackles the problem of large and complex TCBs of current approaches for application-level security policy enforcement. We propose \AppSPEAR{}~-- a security architecture tailored to application-level security policy enforcement. By isolating \AppSPEAR{} components from untrusted application logic and by applying isolation between \AppSPEAR{} components using isolation mechanisms on different hardware/""software-""levels, different implementation variants enable a fine-grained balancing of rigor regarding the reference monitor principles, on the one hand, as well as isolation/""communication costs, on the other hand, and thus, adjusting an \AppSPEAR{} implementation to specific application requirements.

The developer framework supports the implementation by enabling the simple configuration of variants and mechanisms, and the transparent isolation-boundary-crossing communication between the architecture components via proxy objects. The practical evaluation shows that the expected runtime overhead of using TEE/enclave-based isolation can be significantly reduced by using caching and asynchronous enclave calls. While considerably reducing TCB implementation size and complexity compared to conventional mechanisms such as process-based isolation, SGX enables a trusted enforcement of application-specific security policies in \AppSPEAR{} implementations.

Ongoing work focuses on two main areas:
\begin{ilenum}
	\item We are investigating approaches for increasing memory safety of \AppSPEAR{} implementations. The first step is being taken by additionally implementing our developer framework in Rust/Rust-SGX \cite{Wang19a}.
	\item We are further extending developer support: By enabling an automated, compiler-supported code generation of policy and policy runtime environments (\AppSPEAR{} TPS component) from modeled policy representations in domain-specific specification languages \cite{Amthor20a}, we aim to consequently embed our approach into the process of model-based security policy engineering.
\end{ilenum}



\end{document}